\documentclass[aps,prl,twocolumn,showpacs]{revtex4}

\usepackage{amsmath,amsfonts,amssymb}
\usepackage{graphicx}
\usepackage{color,ulem}

\begin{document}

\title{Soliton acceleration by dispersive radiation: a contribution to rogue waves?}

\author{A.~Demircan}
\affiliation{Invalidenstr.~114, 10115 Berlin, Germany}
\author{Sh.~Amiranashvili}
\author{C. Br\'ee}
\affiliation{Weierstrass Institute for Applied Analysis and
Stochastics, Mohrenstra\ss e 39, 10117 Berlin, Germany}

\author{Ch.~Mahnke}
\author{F.~Mitschke}
\affiliation{Institute for Physics, University of Rostock, Universit\"atsplatz 3, 18055 Rostock, Germany}

\author{G.~Steinmeyer}
\affiliation{ Max Born Institute for Nonlinear Optics and Short Pulse Spectroscopy, Max-Born-Stra\ss e 2A, 12489 Berlin, Germany}
\affiliation{{Optoelectronics Research Centre,
Tampere University of Technology, 33101 Tampere, Finland}}
\date{\today}

\begin{abstract}
Rogue waves are solitary waves with extreme amplitudes, which appear to be a ubiquitous
phenomenon in nonlinear wave propagation, with the requirement for a
nonlinearity being their only unifying characteristics. While many
mechanisms have been demonstrated to explain the appearance of rogue
waves in a specific system, there is no known generic mechanism or
general set of criteria shown to rule their appearance. Presupposing
only the existence of a nonlinear Schr\"odinger-type equation
together with a concave dispersion profile around a zero dispersion wavelength
we demonstrate that solitons may experience acceleration and strong
reshaping due to the interaction with continuum radiation, giving rise to extreme-value
phenomena. The mechanism is independent of the optical Raman effect.
A strong increase of the peak power is accompanied by a mild increase
of the pulse energy and carrier frequency, whereas the photon number
of the soliton remains practically constant.
This reshaping mechanism is particularly robust and may explain the appearance
of rogue waves in a large class of systems.
\end{abstract}


\maketitle

The appearance of waves with extreme amplitude has been investigated
in a large class of physical systems
\cite{Dysthe,Ganshin,Bludov,Ruderman,Stenflo,Solli_nat,Kasparian_ox,Majus_pra}.
Their appearance is most drastically illustrated for the case of
ocean waves \cite{Kharif,Janssen,Onorato}, with waves exceeding the
average wave crest by a factor three or more and causing serious
damage to ocean-going ships. Recently, similar phenomena have also
been reported in optics in the soliton-supporting red tail of
supercontinua (SC) in fibers \cite{Solli_nat}. Substantial progress has
been made in the understanding of the mechanisms behind optical
rogue waves
\cite{Solli_prl08,Dudley_ox08,Taki_ox,Akhmediev_pra}.
Currently, most explanations follow one of two alternatives. One
involves soliton fission and selective Raman shifting of the largest
solitons to the long-wavelength side of the spectrum
\cite{Solli_prl08,Dudley_ox08}. The other builds on the
dynamics of Akhmediev breathers
\cite{Akhmediev_pra} and inelastic
collisions between solitons or breathers. While these explanations
concentrate either on soliton fission or fusion processes, we
demonstrate in the following that a similar
mechanism may exist between solitons and continuum radiation in the
normal dispersion range. We will refer to the latter as dispersive
wave (DW). Suitable conditions provided, this DW can strongly modify a soliton through cross phase modulation (XPM).
Strong reshaping, in particular temporal compression, of the soliton is accompanied by a mild increase of 
its energy and carrier frequency, while the photon number of the soliton remains
practically constant. Decrease of the center wavelength gives rise to acceleration of the soliton by virtue of dispersion. The soliton peak power grows surprisingly and may more than double
in only a few centimeters of propagation.

In the following, we consider the rogue wave formation in the SC generation in a
single-mode photonic crystal fiber with one zero dispersion wavelength (ZDW),
which is similar to the fiber used in \cite{Solli_nat,Genty_pla}. In the
following, we restrict our analysis to the minimum set of optical
effects necessary for the formation of rogue waves. These effects
include dispersion and the Kerr nonlinearity, yet exclude Raman
scattering, cf.~\cite{Genty_pla}. In the fiber geometry, the optical
field is characterized by a single real valued component $E(z,t)$
whereas dependencies perpendicular to the propagation coordinate $z$
are integrated out.
We choose a suitable time period $T$, introduce frequencies $\omega\in2\pi\mathbb{Z}/T$, denote spectral field components by $E_{\omega}(z)$, and following \cite{Amiranashvili_pra} define a complex valued $\mathcal{E}(z,t)$ such that $\mathcal{E}_\omega(z)=E_{\omega}(z)-i\partial_{z}E_{\omega}(z)/|\beta(\omega)|$. Note that $E=\mathop{\mathrm{Re}}[\mathcal{E}]$, the propagation equation for $\mathcal{E}(z,t)$ reads 
\begin{equation}\label{asm}
    i\partial_{z}\mathcal{E}_{\omega}+
    |\beta(\omega)|\mathcal{E}_{\omega}+
    \frac{3\omega^2\chi^{(3)}}{8c^2|\beta(\omega)|}(|\mathcal{E}|^2\mathcal{E})_{\omega}=0.
\end{equation}
Parameters $c$, $\chi^{(3)}$, and $\beta(\omega)$ are the speed of light, the third-order nonlinear susceptibility, and the propagation constant respectively. Equation~\eqref{asm} is subject to the conservation laws
\begin{equation}\label{cLaws}
	I_1=\sum_{\omega}\frac{n(\omega)}{\omega}|\mathcal{E}_{\omega}|^2,
	\qquad
	I_2=\sum_{\omega}n(\omega)|\mathcal{E}_{\omega}|^2
\end{equation}
where $n(\omega)$ is refractive index and $I_{1,2}$ are finite and proportional to the time-averaged photon flux and power respectively \cite{Amiranashvili_pra}. For unidirectional propagation $\mathcal{E}(z,t)$ is identical to analytic signal $\mathcal{E}(z,t)=2\sum_{\omega>0}E_{\omega}(z)e^{-i\omega t}$ and moreover only the positive-frequency part of $|\mathcal{E}|^2\mathcal{E}$ contributes to Eq.~\eqref{asm}. The fiber propagation constant may be obtained by numerical integration of the group delay $\beta_1(\omega)=\beta'(\omega)$ and then approximated following \cite{Pade}. If the slow envelope description with respect to a carrier frequency $\omega_0$ applies, Eq.~\eqref{asm} reduces to the standard nonlinear Schr{\"o}dinger equation \cite{Agrawal} with the nonlinearity parameter $\gamma=(3\omega_0\chi^{(3)})/[4\epsilon_0c^2n^2(\omega_0)A_\text{eff}]$, where $A_\text{eff}$ is the effective fiber area.
\begin{figure}
\includegraphics[width=3.2in]{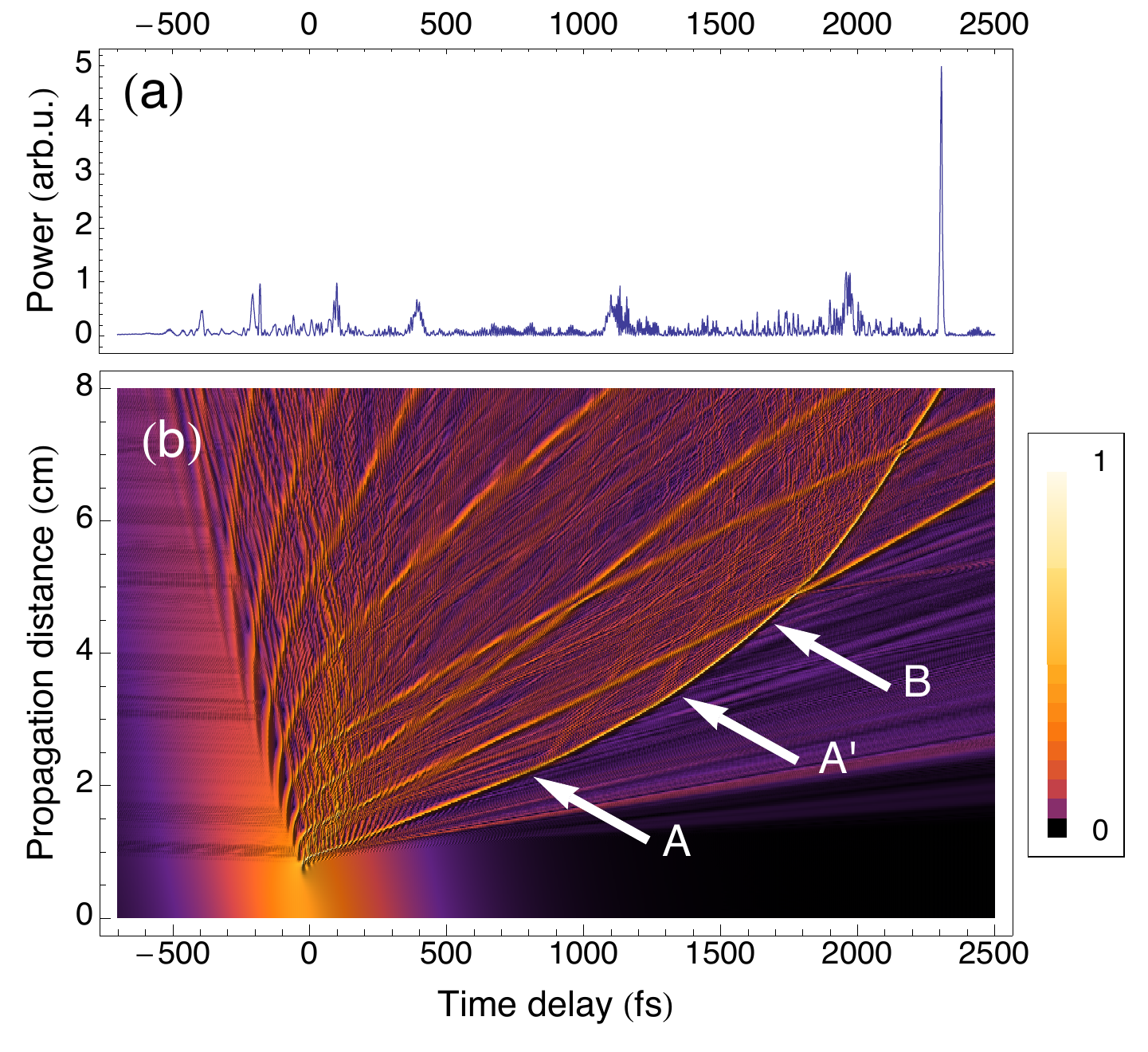}
\caption{Temporal evolution of a higher-order soliton
injected close to the ZDW into the fiber along $z$ for a
typical SC generation process. Note that the calculation 
does not involve the Raman effect.
(a) Final state with a rogue wave exceeding the
average wave crest by more than a factor of three. 
(b) Propagation dynamics of the solitons and non-solitonic radiation, generated by the fission process.}
\label{fig:1}
\end{figure}
Beyond the standard treatment with an envelope approximation, our
approach correctly models nonlinear processes between spectrally
disparate waves, i.e., four-wave mixing processes and XPM between solitons and DWs,
and between individual solitons. These nonlinear processes have been previously
found important for explaining rogue waves in gravity matter waves \cite{Janssen}.

\begin{figure}
\includegraphics[width=3.2in]{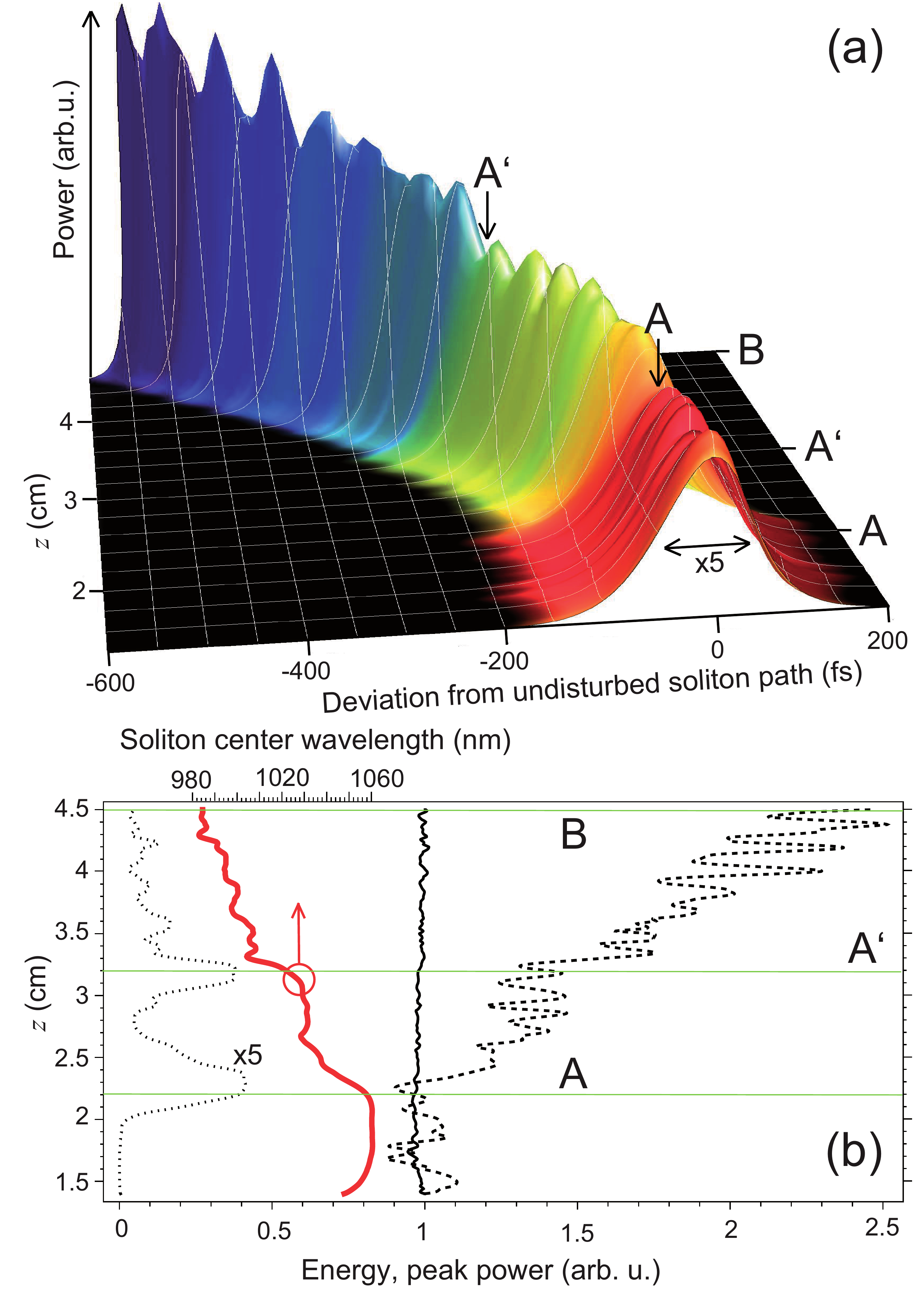}
\caption{(a) Visualization of the soliton propagating from A to B in Fig.~\ref{fig:1}.
Temporal delays are shown relative to the unperturbed propagation of the soliton at $z=1.5$--$2\,$cm.
For clarity, the width of the soliton has been stretched by a factor 5. Color coding visualizes $\lambda_0(z)$,
which changes from 1060 to 985\,nm (red and blue, respectively). (b)
Development of soliton parameters $\lambda_0(z)$ (thick red line),
pulse energy $\propto I_2$ (solid black line), and peak
power $P_0(z)$ (dashed line). Energy content of the dispersive wave
within $\pm1.5\tau$ interval around $t_\ast(z)$ is shown as a dotted line.}
\label{fig:2}
\end{figure}

We launch a hyperbolic secant pulse (center wavelength $897$\,nm, full width at half
maximum $\text{FWHM}=265$ \,fs), corresponding to a higher-order soliton with
soliton number $N\approx 28$ in the anomalous dispersion regime of
the fiber close to the $\text{ZDW}=842$\,nm.
For a nonlinear fiber with $\gamma=0.1\,\text{W}^{-1}\text{m}^{-1}$
this corresponds to a peak power of $19$\,kW,
so that soliton fission is favored \cite{Herrmann_prl}.
These conditions ensure the formation of a
SC with the increase of the initial spectral width by one
to two orders of magnitude \cite{Dudley_rev}. Figure \ref{fig:1}
shows the typical SC evolution in the temporal domain.
With the rather moderate powers in this
example, the effect of the modulational instability can only be observed in the initial
$\approx 0.5$\,cm propagation length before pulses reach the
sub-100\,fs regime \cite{Demircan_apb}. The fundamental solitons
produced in the fission process exhibit durations between 10 and
20\,fs with different peak powers, appearing as pronounced lines
which clearly stand out from the background. 
The fission process also generates DWs in the normal dispersion regime \cite{Dudley_rev}. The further away from
the ZDW solitons are being generated, the slower they will 
propagate  \cite{Demircan_apb}, accumulating delay (Fig.~\ref{fig:1}). As we deliberately
excluded Raman scattering in our analysis,
we might expect that the group velocity of solitons is constant except
in places where isolated soliton-soliton scattering processes occur.
Indeed, inspection of Fig.~\ref{fig:1} reveals several such characteristic
crossings of soliton trajectories in the $t$-$z$ plane. However, it also
reveals that the trajectory of the strongest soliton
does not appear to be ruled by rare isolated scattering events
($\overline{\rm AB}$, Fig.~\ref{fig:1}). The parabolic trajectory of
this soliton is witness of its constant acceleration.

To elucidate the physical mechanisms behind this
peculiar acceleration, we numerically isolated the soliton,
separated it from accompanying continuum radiation, and fitted
the model function $f(t)=P_0 \mathop{\mathrm{sech}}^2 [(t-t_\ast)/t_0]$
to its intensity envelope (Fig.~\ref{fig:2}, FWHM $\tau=1.76t_0$).
Compared to the steady propagation at $z<2.2$\,cm,
Fig.~\ref{fig:2}(a) confirms a deviation of $t_\ast(z)$ from the
initial linear trajectory by $-600$\,fs at point B ($z=4.5$\,cm).
This temporal shift is accompanied by a 4\% change of pulse energy
$\propto I_2$ and by a more than twofold increase
of peak power $P_0(z)$, [solid and dashed curves in
Fig.~\ref{fig:2}(b), respectively]. Pulse duration scales
accordingly from an initial 20\,fs (FWHM) to sub-10 fs at B.
Furthermore, a Fourier analysis indicates that the center wavelength
$\lambda_0(z)$ of the soliton shifts from 1060 to 985\,nm within the
2.3\,cm propagation from A to B, reflecting the according energy transfer.

Rogue waves, subject to non-Gaussian statistics,
have been shown to appear in the fiber SC generation 
both with and without Raman frequency shift in \cite{Genty_pla}. 
The rogue event regarded as an emerging single ``champion'' soliton was linked to multiple collisions between optical solitons.
As in our case there is no other soliton in reach from $A$ to $B$,
this parameter change can only be explained by nonlinear continuum-soliton interaction. 
Namely, for each soliton velocity there is a spectral slice of dispersive
non-solitonic radiation which propagates at nearly identical group velocity.
Group-velocity matching significantly increases the nonlinear interaction length
between continuum and soliton \cite{Dudley_rev,Driben_ox} and may lead
to the reshaping of the latter \cite{Demircan_prl}.
Comparing to the energy of the DW that is in
temporal overlap with the soliton [dotted line in
Fig.~\ref{fig:2}(b)], it is striking that changes of each of
the soliton parameters $t_\ast(z)$, $\lambda_0(z)$, and $P_0(z)$ are
strongly correlated with the strength of the DW, see
positions A and A' marked in Fig.~\ref{fig:2}. It appears
still surprising that a DW with less than $10\%$
of the soliton amplitude can affect its properties so strongly.

\begin{figure}
\includegraphics[width=3.2in]{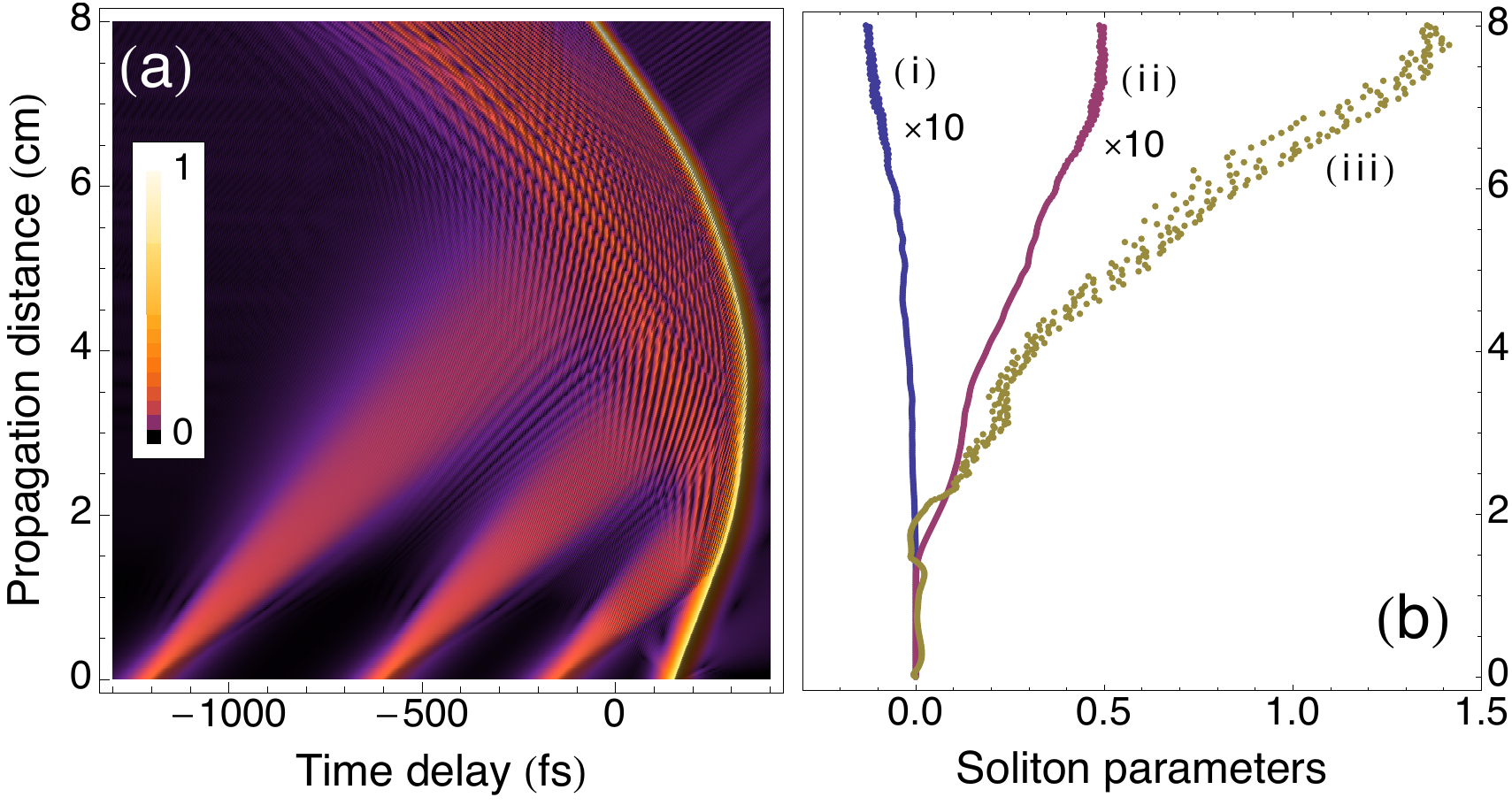}
\caption{(a) Time domain evolution along the fiber representing the
isolated trajectory of a fundamental soliton. Acceleration results
from a cascaded scattering with three DWs. (b) Relative change of soliton parameters:
(i) photon number (see Eq.~\eqref{cLaws} and \cite{Amiranashvili_pra}),
(ii) energy, (iii) peak power.
The deviation of a parameter $\zeta$ is defined as $[\zeta(t)-\zeta(0)]/\zeta(0)$.
For clarity, deviations (i) and (ii) have been stretched by a factor 10.
} \label{fig:3}
\end{figure}

For further investigation of this scenario, we
numerically isolated the primary soliton and selected segments of
the DW in Fig.~\ref{fig:1}(a) right at the onset of
the trajectory curvature, allowing for a deterministic
interpretation of the acceleration process of the soliton uncoupled
from the SC generation process. To this end, we inject
into the fiber a fundamental soliton at $\lambda_s=1030$\,nm of
$26.6$\,fs FWHM duration together with slightly slower propagating
$53.2$\,fs time segments of DWs near the velocity-matched wavelength of
$\lambda_d=614$\,nm.

\begin{figure}
\includegraphics[width=3.2in]{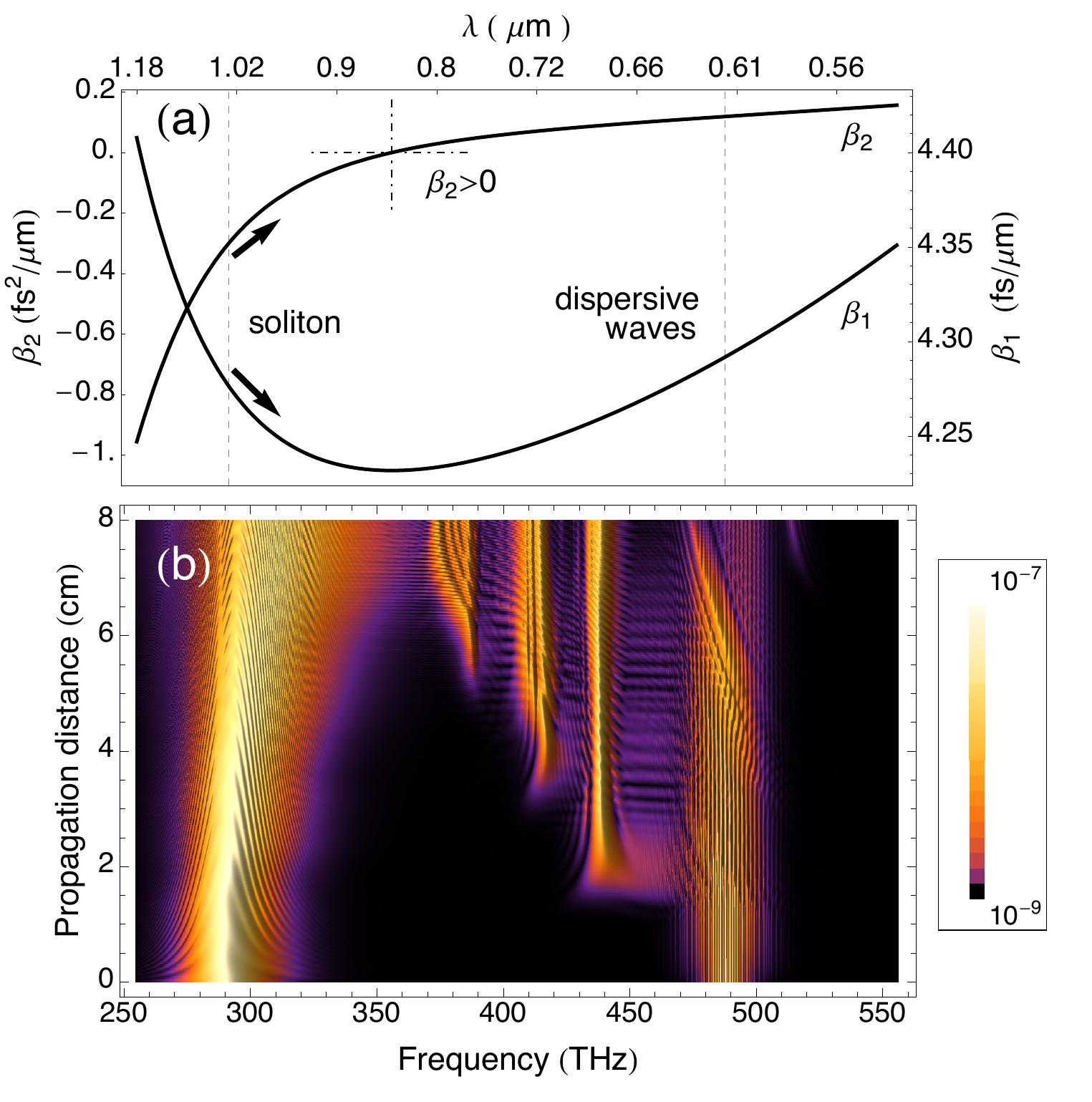}
\caption{(a) Exemplary concave group delay $\beta_1=\beta'(\omega)$
and related group-velocity dispersion $\beta_2=\beta''(\omega)$, with the extracted wavelengths for the fundamental soliton at $\lambda_s=1030$\,nm and a dispersive pulse
at $\lambda_d=614$\,nm (dashed line).
Arrows indicate the induced change of $\beta_2$ and $\beta_1$ for the soliton. (b) Spectral evolution along the fiber representing a cascaded scattering with three DWs
from a soliton.}
\label{fig:4}
\end{figure}

Figure \ref{fig:3}(a) demonstrates an example for a
continuum-soliton scattering process with three DWs in
a suitably chosen reference frame. Each collision leads to a
stepwise acceleration of the soliton, clearly confirming the transfer of energy and the
concomitant gradual increase in peak power [Fig.~\ref{fig:3}(b)] 
as previously seen in Fig.~\ref{fig:2}. As all obscuring
continuum components have been eliminated, the role of the
dispersive radiation can now be seen in much greater
clarity. The DWs initially propagate at slightly lower
group velocity than the trailing soliton so that they eventually collide.
In this collision, the soliton can never pass the DW as would be expected
in a purely linear-optical encounter. Instead, XPM between DW
and soliton causes a frequency shift towards the ZDW
($842$\,nm), decreasing the center wavelength of the soliton and
shifting the DW toward longer wavelength. Comparing to
the underlying dispersion profile [Fig.~\ref{fig:4}(a)], both these
shifts lead to an acceleration for the respective type of radiation,
as is clearly confirmed by the trajectory curvatures in
Fig.~\ref{fig:3}(a).

Under similar conditions, the impenetrability of the
soliton trajectory was referred to as an optical event horizon for
the DW \cite{Hawking,Demircan_prl}. Its origin lies in
a nonlinearly induced increase of the group velocity caused by the
leading edge of the soliton. The only way for the DW to
escape from the event horizon is a shift towards the ZDW, i.e., both
types of radiation therefore experience a strongly enhanced
effective XPM. These processes are completely elastic, causing a
mutual shift of optical frequencies but never transferring photons
from the normal dispersion regime into the soliton regime or vice versa.
With the photon number of the soliton practically conserved,
the soliton blue shift accordingly causes a mild increase of its energy [Fig.~\ref{fig:3}(b)].
The shifted soliton also experiences a considerably smaller $\beta_2$ [Fig.~\ref{fig:4}(a,b)]. 
Now consider that the energy of a fundamental soliton can be expressed through $P_0$
and $\beta_2$ as $E=2\sqrt{P_0|\beta_2|/\gamma}$. Obviously, the decrease of $\beta_2$
cannot be compensated by a reduction of $E$ because $E$ also grows. 
As $\gamma$ does not vary appreciably, consequently, $P_0$
is forced to grow massively. This clearly explains our observations.
We repeated these simulations with several segments
of continuum to prove that even higher soliton peak
powers can be achieved with segments of the continuum containing more
energy.

The nature of the newly observed continuum-soliton
scattering processes is markedly different from soliton-soliton
scattering. As continuum radiation quickly disperses, there will
always be temporal slices of the DW that effectively
copropagate with a given soliton, making mutual extended interaction
much more likely than the appearance of soliton-soliton processes.
We therefore suggest that this mechanism contributes to the dramatic
amplitude increases seen in experimental work \cite{Solli_nat}.

The mechanism proposed here does not
presuppose any special nonlinear effects that are unique to optical
systems. In comparison to previously discussed mechanisms of rogue
wave formation, our approach essentially only presupposes a
nonlinear Schr{\"o}dinger type scenario, with a reactive
nonlinearity and a concave dispersion profile, the
latter enabling copropagation of radiation with opposite signs of
dispersion with equal group velocity. These conditions are met in a
variety of systems, e.g., for gravity-capillary waves \cite{Lamb}.
Our explanation is therefore immediately
applicable to a much wider class of physical systems. Consequently,
we believe that the previously disregarded scattering of DWs
off solitons opens a new perspective on the fascinating
appearance of extreme-value wave phenomena.

The following support is gratefully acknowledged:
Sh.~A. by the DFG Research Center
MATHEON (project D\,14), C.~M.~and F.~M.~by DFG, and G.~S.~
by the Academy of Finland (project grant 128844).


\begin{thebibliography}{99}
\bibitem{Dysthe} K.~Dysthe, H.~E.~Krogstad, and P.~M\"uller, Annu.~Rev.~Fluid Mech.~{\bf 40}, 287 (2008).
\bibitem{Ganshin} A.~N.~Ganshin {\it et al.}, Phys.~Rev.~Lett.~{\bf 101}, 065303 (2008).
\bibitem{Bludov} Yu.~V.~Bludov, V.~V.~Konotop, and N.~Akhmediev, Phys.~Rev.~A {\bf 80}, 033610 (2009).
\bibitem{Ruderman} M.~S.~Ruderman, Eur.~Phys.~J.~Special Topics {\bf 185}, 57 (2010).
\bibitem{Stenflo} L.~Stenflo and M.~Marklund, J.~Plasma~Phys.~{\bf 76}, 293 (2010).
\bibitem{Kasparian_ox} J.~Kasparian {\it et al.}, Opt.~Express {\bf 17}, 12070 (2009).
\bibitem{Majus_pra} D.~Majus  {\it et al.}, Phys.~Rev.~A {\bf 83}, 025802 (2011).
\bibitem{Solli_nat} D.~R.~Solli {\it et al.}, Nature {\bf 450}, 1054 (2007).
\bibitem{Kharif} C.~Kharif and E.~Pelinovsky, Eur.~J.~Mech.~{\bf 22}, 603 (2003).
\bibitem{Janssen} P.~A.~E.~M.~Jannsen, J. Phys. Oceanography {\bf 33}, 863 (2003).
\bibitem{Onorato} M.~Onorato {\it et al.}, Phys.~Rev.~Lett.~{\bf 86}, 5831 (2001).
\bibitem{Taki_ox} A.~Mussot {\it et al.}, Opt.~Exp. {\bf 17}, 1502 (2009); M.~Taki  {\it et al.}, Phys.~Lett.~A {\bf 374}, 691 (2010).
\bibitem{Solli_prl08} D.~R.~Solli, C.~Ropers, and B. Jalali, Phys.~Rev.~Lett {\bf 101}, 233902 (2008).
\bibitem{Dudley_ox08} J.~M.~Dudley, G.~Genty, B.~J.~Eggleton, Opt.~Exp. {\bf 16}, 3644 (2008); M.~Ekintalo, G.~Genty, and J.~M.~Dudley, Opt.~Lett. {\bf 35}, 658 (2010); G.~Genty, J.~M.~Dudley, B.~J.~Eggleton, Appl.~Phys.~B {\bf 94}, 187 (2009).
\bibitem{Akhmediev_pra} N.~Akhmediev, J.~M.~Soto-Crespo, and A.~Ankiewicz, Phys. Rev. A {\bf 80}, 043818 (2009);  N.~Akhmediev,  A.~Ankiewicz, and M.~Taki, Phys.~Lett. A {\bf 373}, 675 (2009); B.~Kibler {\it et al.}, Nature Physics {\bf 6}, 790 (2010).
\bibitem{Genty_pla} G.~Genty {\it et al.}, Phys.~Lett.~A {\bf 374} 989 (2010).
\bibitem{Amiranashvili_pra} Sh.~Amiranashvili, A.~Demircan, Phys.~Rev.~A {\bf 82}, 013812 (2010).
\bibitem{Pade} Sh.~Amiranashvili, U.~Bandelow, A.~Mielke, Opt. Commun. {\bf 283}, 480 (2009).
\bibitem{Agrawal} G.~Agrawal, {\it Nonlinear Fiber Optics} (Academic Press, San Diego, 2001).
\bibitem{Herrmann_prl} J.~Herrmann {\it et al.}, Phys. Rev. Lett. {\bf 88}, 173901 (2002).
\bibitem{Dudley_rev} J.~M.~Dudley, G.~Genty, S.~Coen, Rev. Mod. Phys. {\bf 78}, 1135 (2006).
\bibitem{Demircan_apb} A.~Demircan and U. Bandelow, Appl.~Phys.~B {\bf 86}, 31 (2007); Opt.~Comm. {\bf 244}, 181 (2005).
\bibitem{Driben_ox} R.~Driben, F.~Mitschke, and N. Zhavoronkov, Opt.~Exp.~{\bf 18}, 25993 (2010)
\bibitem{Philbin_sc} T.~G.~Philbin {\it et al.},, Science {\bf 319}, 1367 (2008).
\bibitem{Hawking} F. Belgiorno {\it et al.}, Phys. Rev. Lett. {\bf 105}, 203901 (2010).
\bibitem{Demircan_prl} A.~Demircan, Sh.~Amiranashvili, G.~Steinmeyer, Phys.~Rev.~Lett. {\bf 106}, 163901(2011).
\bibitem{Lamb} H.~Lamb, {\it Hydrodynamics} (6th ed.), Cambridge University Press (1994).



\end{thebibliography}
\end{document}